\documentclass[draftcls, onecolumn,12pt]{IEEEtran}

\usepackage{amsmath}
\usepackage{cases}
\usepackage{amsfonts}
\usepackage{amssymb}
\usepackage{boxedminipage}
\usepackage{graphicx}
\usepackage[usenames]{color}
\usepackage{array,color}
\usepackage{colortbl}

\newcommand{\D}{\displaystyle}
\newcommand{\DF}[2]{\frac{\D#1}{\D#2}}

\begin{document}
\title{Throughput Improvement and Its Tradeoff with The Queuing Delay in the Diamond Relay
Networks}

\author{Qing~Wang,
        Pingyi~Fan,~\IEEEmembership{Senior Member,~IEEE,}
        and Khaled~Ben~Letaief,~\IEEEmembership{Fellow,~IEEE}\\
\{qing-wang06@mails,~fpy@mail\}.thu.edu.cn, ~eekhaled@ece.ust.hk}

 \maketitle
\baselineskip=18pt
\begin{abstract}
\baselineskip 18pt
 Diamond relay channel model, as a basic transmission model,  has recently been
attracting considerable attention in wireless Ad Hoc networks. Node
cooperation and opportunistic scheduling scheme are two important
techniques to improve the performance in wireless scenarios. In the
paper we consider such a problem how to efficiently combine
opportunistic scheduling and cooperative modes in the Rayleigh
fading scenarios. To do so, we first compare the throughput of SRP
(Spatial Reused Pattern) and AFP (Amplify Forwarding Pattern) in the
half-duplex case with the assumption that channel side information
is known to all and then come up with a new scheduling scheme. It
will that that only switching between SRP and AFP simply does little
help to obtain an expected improvement because SRP is always
superior to AFP on average due to its efficient spatial reuse. To
improve the throughput further, we put forward a new processing
strategy in which buffers are employed at both relays in SRP mode.
By efficiently utilizing the links with relatively higher gains, the
throughput can be greatly improved at a cost of queuing delay.
Furthermore, we shall quantitatively evaluate the queuing delay and
the tradeoff between the throughput and the additional queuing
delay. Finally, to realize our developed strategy and make sure it
always run at stable status, we present two criteria and an
algorithm on the selection and adjustment of the switching
thresholds .

\end{abstract}



\section{Introduction}
In recent years, motivated by higher quality-demanded applications
in the wireless Ad Hoc networks, cooperation among nodes is
considered to be more useful over various relay network models.
Early in 1970s, a classical three-node relay channel comprised of
one source, one destination and only one relay was first introduced
by van de Meulen in [3]. Then, in reference [4] and [5], the cut-set
bound and the achievable rates with power allocation were studied
for the half-duplex case in this three-node network. Especially,
Reference [5] presented some lower and upper bounds on outage
capacity of three-node network. The analysis of
diversity-multiplexing tradeoff for the three-node network in the
half-duplex case was given in [6]. Besides, systems using multiple
relays were also studied. In [7], both the achievable rates and the
upper bound of capacity were studied for the diamond relay network
in the half-duplex case. The two-relay model (i.e. the Diamond Relay
Networks) operating in the full-duplex case were studied in [8]. In
[9], they analyzed a model in which cooperative communication
proceeds in a parallel relay network where the exogenous arrival of
packets and the FIFO (First In First Out) queueing system are
introduced. Multiple relays using orthogonal channels were analyzed
in [10]. Besides, relay networks with and without delay were
discussed in [12] and [13], respectively. Reference [14]-[19]
presented analysis of capacity and delay tradeoff in the networks
comprised of many random prelocated nodes. These works showed that a
cooperative gains can be obtained in distributed wireless networks
if nodes can help each other to relay information. This motivated us
to study the wireless relay networks further, especially for some
classical topologies and cooperative modes.

Referring to the wireless relay networks, the three-node network has
been a very hot topic in the research area of the cooperative
network since 1970s due to its classical and representative
topology. In the late ninety's, B.Schein and R.Gallager proposed
another kind of relay network, the Gaussian parallel relay network
[8], in which the diamond relay network was first introduced
implicitly. The reasons that diamond relay networks were considered
include two facets: Firstly, this kind of topology is relatively
easy traceable in theory and has more freedom than three-node
network model. Secondly, this model can be used in some wireless
scenarios, where a sender is convenient to select a few neighbors.
Fig. 1 shows one application scenario where the black and square
parts represent the obstacles blocking radio signals seriously,
e.g., skyscrapers in business district. In such a scenarios, if node
A wants to transmit to node D at a higher rate, the diamond relay
network comprised of node A,B,C and D can do a great help.


Based on different topologies of the wireless relay networks
[3]-[8],[10], several cooperative modes were introduced
consequently, such as SRP (Spatial Reused Pattern) and AFP (Amplify
Forwarding Pattern) in [7]. In the wireless time-varying and fading
scenarios, different cooperative modes may have big difference in
term of reliable throughput where opportunistic scheduling may play
a key role in improving the throughput. Motivated by this, we
firstly review the two different cooperative modes, namely SRP and
AFP, and present their corresponding capacities in the diamond relay
network, shown in Fig. 2, in the half-duplex case. In the
discussion, relay nodes adopt two relay schemes, namely, the
amplify-and-forward (AF - relay node simply amplifies the signals
received from source and forwards to destination) and
decode-and-forward (DF - relay node decodes the information received
from  source, re-encodes and forwards  to  destination). In the
sequel, AF relay scheme is referred to AFP (Amplify Forwarding
Pattern) mode while DF relay scheme is referred to SRP(Spatial
Reused Pattern) mode. Furthermore, we compare the throughput of the
two cooperative modes and find that there exists a big difference
between the performances of the SRP and AFP mode. In some cases, SRP
has a larger throughput than  AFP. Otherwise, it has smaller
throughput than AFP.  Thus, our first contribution is that we
analyze the efficiency of opportunistic scheduling and put forward a
hybrid relay scheme with switching between SRP and AFP so that it
can be adapted to the channel variation.


Although the opportunistic scheduling is usually considered to be
effective, it will be shown that combining SRP and AFP in a simple
way can not obtain an expected improvement. This is because SRP is
always superior to AFP on average due to the efficient spatial
reuse. In previous works [3]-[8], using Max-Flow Min-Cut theorem, it
can be seen that the whole performance is always reduced greatly by
some bottleneck links usually caused by the fading and time-vary
characteristics of the wireless channels. Therefore, using buffers
at nodes maybe help to efficiently employ some channels in rather
good conditions. In this way, it will improve the whole performance
of the throughput in the wireless networks at the cost of some
additional queuing delay. Motivated by this, we shall put forward a
new processing strategy in which buffers are employed at both relays
in SRP mode. Its basic idea is that in  diamond relay channels,
sometimes there is only one or two links in very good conditions
while the others in bad ones. In this case, buffers can help relays
efficiently utilizing the links in good conditions so that the
throughput is greatly improved at some cost of queuing delay. One
important contribution of this work is that we quantitatively
evaluate the queuing delay and discuss its tradeoff with the
throughput where two criteria are considered to characterize the
thresholds for separating pretty good or rather bad condition of
channels and make sure the network run in stable status. In
addition, an adjust algorithm is also given.

For simplicity, finite state channel model is used in which the
received SNR is partitioned into $N$ levels. The status of the four
links are assumed to be independent and each link operates in a
memoryless mode. G/G/1 queueing system is used to model the relay
node with buffers and the associated two links connecting to it.
Besides, a continuous traffic model is considered in which the
source always has information to send. Then one can use Marshall's
theory to solve it. Note that here we consider the source always has
information to send, which will help us to get more insights on the
maximum achievable throughput of this diamond network. Thus, other
traffic models, such as Bernoulli or Poisson traffic model, will not
be discussed here. Another point should be mentioned is that due to
the IC design improvement, buffers with enough capacity are becoming
much cheap and with low cost, so the delay is a more rigorous factor
influencing the performance of the networks. Therefore we emphasize
the average queuing time without limitation of the buffer sizes,
referred to delay in this paper.

The rest of this paper is organized as follows: In Section
\ref{model1}, we introduce the system model. In Section
\ref{model2}, we firstly compare the performance of SRP and AFP
scheme,then we give an opportunistic scheduling scheme, namely an
hybrid scheme of combining SRP and AFP modes. In Section
\ref{model3}, we shall propose a new processing strategy in which
buffers are employed at the relay nodes in SRP mode. We present two
criteria to characterize the thresholds for realizing tradeoff
between the throughput and queuing delay and make sure the network
run in stable status. Some simulation results are given in Section
\ref{model4}. Finally, We present the conclusion in section
\ref{model5}.

\section{System Model}
\label{model1}

\subsection{Diamond Relay Channel and Two Time Sharing Patterns: SRP
and AFP}

The discussion of diamond relay channel was first considered in
Schein and Gallager's work in a full-duplex case [8]. Recently,
Feng, etc, studied this kind of networks in the half-duplex case in
[7], shown in Fig. 2.  It is comprised of four nodes including a
source $N_s$, two relays $N_1$ and $N_2$ and a destination $N_d$. It
is assumed that all four nodes operate in half-duplex mode, and that
the destination can not communicate with the source directly and the
two relay nodes will not interfere with each other [7]. Now we first
review the two cooperative modes, namely SRP and AFP, as follow.\\
\textbf{SRP: }A time block, $T$, is divided into two stages.
\begin{enumerate}
\item {\em Stage 1}: In the first $\lambda T$ slots, source node $N_s$ and relay node
$N_2$ transmit signals while relay node $N_1$ and destination node
$N_d$ are in receiving status. $\lambda \in [0,1]$ is time sharing
parameter.
\item {\em Stage 2}: In the remaining $(1-\lambda)T$ slots, source
node $N_s$ and relay node $N_1$  transmit signals while relay node
$N_2$ and destination node $N_d$ are in receiving status.
\end{enumerate}
\textbf{AFP: }A time block is also divided into two stages.
\begin{enumerate}
\item {\em Stage 1}: In the first $\lambda T$ slots, source node
$N_s$ transmit signals while both relays $N_1$ and $N_2$ are in
receiving status. In this case, the destination can not hear the
signal.
\item {\em Stage 2}: In the next $(1-\lambda)T$ slots, both relays
$N_1$ and $N_2$  forward the received signals in the first $\lambda
T$ time slots while the destination is in receiving status and
source node keeps silent.
\end{enumerate}

Let $X_i(t)$ and $Y_j(t)$ denote the signal sent by node $i$  and
that received by node $j$, respectively. Then the equivalent
baseband signals of the two cooperative modes are given by,
respectively,
\begin{enumerate}
\item In SRP:\\
In the first $\lambda T$
slots:$\hspace{16mm}Y_1(t)=g_{s1}X_s(t)+Z_{s1}(t)\,\,,\,\,
Y_d(t)=g_{2d}X_2(t)+Z_{2d}(t)$ \\
In the remained $(1-\lambda)T$
slots:$\hspace{5mm}Y_2(t)=g_{s2}X_s(t)+Z_{s2}(t)\,\,,\,\,
Y_d(t)=g_{1d}X_1(t)+Z_{1d}(t)$
\item In AFP:\\
In the first $\lambda T$
slots:$\hspace{16mm}Y_1(t)=g_{s1}X_s(t)+Z_{s1}(t)\,\,,\,\,
Y_2(t)=g_{s2}X_s(t)+Z_{s2}(t)$\\
In the remained $(1-\lambda)T$
slots:$\hspace{5mm}Y_d(t)=g_{1d}X_1(t)+g_{2d}X_2(t)+Z_{12d}(t)$
\end{enumerate}
where $\{Z_{s1}(t),Z_{s2}(t),Z_{1d}(t),Z_{2d}(t),Z_{12d}(t)\}$ are
independent and identically distributed (i.i.d) circular Gaussian
random variables. $g_{ij}$ is the gain factor of the link from node
$i$ to node $j$. In the discussion, all the nodes $i\in
\{N_s,N_1,N_2,N_d\}$ are power limited. Their maximum transmitted
are assumed to be the same, denoted as {\em $P_c$}. Here the channel
side information is used by receivers feedback few bits reflecting
the link conditions. The bit number is relatively little compared to
the data packets.  For instance, one relay can first estimate the
status of the two links associated to it and then feedback to the
source and the destination. The source will inform the status of
this link to another relay in next time slot by adding a overhead in
its traffic massage. This process renews only once at the beginning
of each time block according to the block channel fading. In this
way, all the nodes could obtain the channel side information and
cooperate in this time block, $T$, which consists of many time
slots.

\subsection{Finite State Fading Channel}
To effectively analyze the fading and time-variant characteristics
of channels, a finite state fading model is built by partitioning
the instantaneous received SNR into N levels. Some partition methods
, such as the equal-probability partition method [20], the optimum
Minimum Mean-Square Error (MMSE) Lloyd-Max quantification method
etc. have been employed previously.  For simplicity, we shall adopt
an equal-probability partition and use its mean value of SNR to
represent the exact SNRs in each interval [20] here.  Note that it
will be consistent with the real case as the partition level is
large enough.

Let $\pi_i$  $ (1 \leq i \leq N)$ denote the probability of link
state $i$ and $B(i)$ $(1 \leq i \leq {N-1})$ denote the partition
boundary of each state, which is determined by
\begin{eqnarray}
 & & \pi_1 = \pi_2 = \cdots = \pi_{N-1} = \pi_N = \frac1N\\
 & & \int_{B(i)}^{B(i+1)}f(x)\,\mathrm{d}x =\frac1N ;\,
 i\in[1,N-2],\,\int_0^{B(1)}f(x)\,\mathrm{d}x =\frac1N\,\,,\,\, \int_{B(N-1)}^{+\infty}f(x)\,\mathrm{d}x =\frac1N
\end{eqnarray}
where $f(x)=\lambda e^{-\lambda x}$ is the {\em $p.d.f.$} of the
received SNR over a Rayleigh channel and $\lambda$ is the reciprocal
of the average received $\overline{SNR}$. The mean value of SNRs in
each interval, $\overline{SNR_i}$,$(1 \leq i \leq {N-1})$, is given
by

\begin{eqnarray}
\overline{SNR_i}=\frac{\int_{B(i)}^{B(i+1)} x \cdot
f(x)\,\mathrm{d}x}{\int_{B(i)}^{B(i+1)}f(x)\,\mathrm{d}x}=N\int_{B(i)}^{B(i+1)}
x f(x)\,\mathrm{d}x\,\,,\,\,
\overline{SNR_N}=N\int_{B(N-1)}^{\infty} x f(x)\,\mathrm{d}x
\end{eqnarray}

The maximum average rate $C_i$ of the link at the state level {\em
i} is given by $ C_i=\frac12 \log(1+\overline{SNR_i})$.

 In the paper, we assume that the maximum rate can be approximately achieved by
some effective pseudo-random channel coding schemes and  that the
destination node can receive the signals from the two relays
coherently in the AFP model. Thus, we use the corresponding capacity
to approximately trace the maximum achievable transmitting rate. For
simplicity,  let $C$ denote the maximum achievable rate in the
sequel.

\subsection{Block Rayleigh Fading Model}
It is assumed that the channels endure block fading, which means
that the received SNR in one time block $T$ is a constant, but it
may vary from block to block obeying the exponential distribution,
which is corresponding to Rayleigh fading. Furthermore, the states
of the four links are assumed to be independent and for each link
its SNR varies according to a memoryless mode.

\subsection{Marshall's Queueing Theory}
For G/G/1 model, Marshall's theorem on the estimation of the average
queuing time is reviewed here [11].

\textbf{Theorem $2.1$:} For all G/G/1 queues with $\rho < 1$, we
have
\begin{equation}
E(W)=\frac{\lambda^2(\sigma_a^2+\sigma_b^2)+(1-\rho)^2}{2\lambda(1-\rho)}-\frac{\upsilon_h^{(2)}}{2\upsilon_h}
\end{equation}
where $a$ and $b$ denote the arrival interval and the service time,
respectively. $\sigma_a^2$ and $\sigma_b^2$ denote the corresponding
variances of them, respectively. $\lambda$ is the average traffic
arrival rate and $\rho$ is the traffic intensity of the system.
$\upsilon_h$ and $\upsilon_h^{(2)}$ are the first and second order
moments of the idle period $h$ of the system.

If the inequality $\frac{\upsilon_h^{(2)}}{\upsilon_h} \ge
\frac1\lambda (1-\rho)$ holds, the upper bound of the waiting time
is given by
\begin{eqnarray}
E(W)\le\frac{\lambda^2(\sigma_a^2+\sigma_b^2)+(1-\rho)^2}{2\lambda(1-\rho)}-\frac{1}{2\lambda}(1-\rho)
=\frac{\lambda(\sigma_a^2+\sigma_b^2)}{2(1-\rho)}
\end{eqnarray}
Note that the inequality $(5)$ becomes an equality when $\rho$
approaches $1$.

\section{Opportunistic Scheduling Scheme Combining SRP and AFP}
\label{model2}

To compare the performance of cooperative modes, SRP and AFP, we
first analyze the capacity of the SRP and AFP. For convenience,
several symbols are defined first.\\
\textbf{{\em (1)} $G_{ij}$} is defined as $G_{ij}=\frac{P_c
||g_{ij}||^2}{\sigma^2}$ and $C_{ij}=\frac12 \log(1+G_{ij})$\\
\textbf{{\em (2)} $x,y\in \mathbb{R}^+$} are defined as follows,
respectively. $x=C_{1d}C_{2d}-C_{s1}C_{s2}$,
$y=C_{s2}C_{1d}-C_{s1}C_{2d}$.

\subsection{Capacity of SRP and AFP Modes}
The SRP mode is a 2-hop strategy in which relay nodes decode their
recdeived information first before re-transmitting to the
destination. It is an efficient cooperative scheme for the diamond
relay model due to the full spatial reuse. In the AFP mode, both
relay nodes just amplify the  signals received in the first half of
{\em T} and re-transmit it in the next one. At each relay node, the
signal is multiplied with a constant and the amplified signals from
the two relay nodes are coherently added up at the destination if
the timing synchronization and carrier recovery are perfect. In both
modes, no buffer is used by the relay nodes.  The maximum achievable
transmission rate is based on the capacities between the links
associated to the relay nodes.

\textbf{Theorem $3.1$:}[7,Theorem4.1 and 5.2]  {\em (i)} In SRP
mode, $(C_{ij}, \hspace{2mm} i \in \{s,1,2\},j \in \{1,2,d\})$
denotes the capacity of the link from node $i$ to node $j$.
Transmitting rate of the link $N_s - N_1 - N_d$ is denoted as $C_1$
and the one of the link $N_s - N_2 - N_d$ is denoted as $C_2$. The
capacity of the SRP mode denoted as $C_{SR}$, which also represents
the maximum achievable rate, is given by
\begin{eqnarray}
C_{SR}=\max_{\lambda_1,\lambda_2}\{C_1 + C_2\}&=&\max_{\lambda_1,\lambda_2}\{(\lambda_1 C_{s1}+\min_{\lambda_1}\{\lambda_1 C_{2d},(1-\lambda_1) C_{s2}\}), \nonumber \\
&&\hspace{3mm}(\lambda_2 C_{2d}+\min_{\lambda_2}\{\lambda_2
C_{1d},(1-\lambda_2) C_{s1}\})\}
\end{eqnarray}
in which $\lambda_1=C_{1d}/(C_{s1}+C_{1d})$ and
$\lambda_2=C_{s2}/(C_{s2}+C_{2d})$.

In addition, the link-state space can be divided into four different
subspace according to the following conditions (7) (8) (9) and (10).
The explicit expression of capacity for the SRP mode in each case is
given as follow.
\begin{eqnarray}
&If&(x\ge 0\bigcap y\ge 0 \bigcap yC_{s1}\ge xC_{s2}) \bigcup (x\ge 0\bigcap y< 0 \bigcap |y|C_{1d}\ge xC_{s2})\nonumber\\
&&\Longrightarrow C_{SR}=\frac{C_{s1}(C_{1d}+C_{s2})}{C_{1d}+C_{s1}} \\
&If&(y\ge 0\bigcap x\ge 0 \bigcap xC_{s2}> yC_{s1}) \bigcup (y\ge 0\bigcap x< 0 \bigcap |x|C_{2d}> yC_{s1})\nonumber\\
&&\Longrightarrow C_{SR}=\frac{C_{2d}(C_{s1}+C_{s2})}{C_{2d}+C_{s2}} \\
&If&(x< 0\bigcap y\ge 0 \bigcap |x|C_{2d}\le yC_{s1}) \bigcup (x< 0\bigcap y< 0 \bigcap |y|C_{1d}\ge |x|C_{2d})\nonumber\\
&&\Longrightarrow C_{SR}=\frac{C_{1d}(C_{s1}+C_{2d})}{C_{1d}+C_{s1}} \\
&If&(y< 0\bigcap x\ge 0 \bigcap |y|C_{1d}< xC_{s2}) \bigcup (y< 0\bigcap x< 0 \bigcap |y|C_{1d}< |x|C_{2d})\nonumber\\
&&\Longrightarrow C_{SR}=\frac{C_{s2}(C_{1d}+C_{2d})}{C_{2d}+C_{s2}}
\end{eqnarray}
{\em (ii)} In AFP mode, parameter $\alpha$ and $\beta$ denote the
amplified factors at relay node $N_1$ and $N_2$, respectively. Since
the signals are received coherently, the maximum achievable rate is
\begin{eqnarray}
C_{AF}&=&\max_{\alpha,\beta}\{\frac12 \cdot \frac12\log
(1+\frac{(\alpha ||g_{s1}||+\beta ||g_{s2}||)^2}{\alpha^2+\beta^2+1}
\frac{P_c}{\sigma^2}) \} \nonumber \\
&\leq&\max_{\alpha,\beta}\{\frac12 \cdot \frac12\log (1+
\frac{(\alpha^2+\beta^2)(||g_{s1}||^2+
||g_{s2}||^2)}{\alpha^2+\beta^2+1} \frac{P_c}{\sigma^2}) \}
\end{eqnarray}
\begin{eqnarray}
&s.t.&\alpha^2(1+\frac{||g_{s1}||P_c}{\sigma^2})\leq
\frac{||g_{1d}||P_c}{\sigma^2}\,\,,\,\,
\beta^2(1+\frac{||g_{s2}||P_c}{\sigma^2})\leq
\frac{||g_{2d}||P_c}{\sigma^2}.\nonumber
\end{eqnarray}
The first factor $\frac12$ is due to the equal time-sharing and the
first inequality becomes equality when $\alpha / \beta =
{||g_{s1}||}/{||g_{s2}||}$ holds.

Based on the theorem above, one can compare the maximum achievable
rate of the two cooperative modes. However, Theorem 3.1 only
consider the case where all the link capacity are fixed. If all the
links are time varying, it is possible for us to select an effective
processing mode adapted to the variation of the links so that we can
get larger throughput by using buffers at relay nodes, which will
shown later. Numerical results in Fig. 5 indicate that the upper
bound of $C_{AF}$ in Eqn.(11) is much smaller than $C_{SR}$ on
average. In addition, a general form of $C_{AF}$ will be given in
Appendix (B).

\subsection{Comparison Between SRP and AFP and An Opportunistic Scheduling Scheme}
Using Eqn.(6) and Eqn.(11), we can divide the link-state space
spanned by the four channel gain factors,
$\{g_{s1},g_{s2},g_{1d},g_{2d}\}$, into eight different subspace. In
each subspace, the capacity for both SRP and AFP modes are
completely determined. Now, let us see a special case on AFP.

\textbf{Theorem $3.2$:} In AFP mode, the link-state space is divided
into two different subspace by the following conditions, where
$\alpha / \beta = {||g_{s1}||}/{||g_{s2}||}$ is satisfied. The
corresponding explicit expression of $C_{AF}$ in each case is given
by
\begin{eqnarray}
&If&\,\,\, \frac{G_{2d}}{G_{s2}(G_{s2}+1)} \geq
\frac{G_{1d}}{G_{s1}(G_{s1}+1)} \nonumber \\
&C_{AF}&=\frac12\cdot \frac12 \log (1+\frac{P_c}{\sigma^2}\cdot
\frac{||g_{1d}||^2 (||g_{s1}||^2+||g_{s2}||^2)^2}{||g_{1d}||^2
(||g_{s1}||^2+||g_{s2}||^2)+||g_{s1}||^2
(||g_{s1}||^2+\frac{\sigma^2}{P_c})})
\end{eqnarray}
\begin{eqnarray}
&If&\,\,\, \frac{G_{2d}}{G_{s2}(G_{s2}+1)} <
\frac{G_{1d}}{G_{s1}(G_{s1}+1)} \nonumber \\
&C_{AF}&=\frac12\cdot \frac12 \log (1+\frac{P_c}{\sigma^2}\cdot
\frac{||g_{2d}||^2 (||g_{s1}||^2+||g_{s2}||^2)^2}{||g_{2d}||^2
(||g_{s1}||^2+||g_{s2}||^2)+||g_{s2}||^2
(||g_{s2}||^2+\frac{\sigma^2}{P_c})})
\end{eqnarray}

The proof is given in Appendix (A), while a general expression of
$C_{AF}$ is given in Appendix (B).

Based on $Theorem\,\,3.1$ and $Theorem\,\,3.2$, the maximum average
achievable rate can be explicitly presented in theory for all the
different subspace so that the comparison between them becomes
traceable.

The whole link state space spanned by
$\{g_{s1},g_{s2},g_{1d},g_{2d}\}$ is divided into eight different
subspace and they are given by
\begin{eqnarray*}
subset (a)=\{g_{s1},g_{s2},g_{1d},g_{2d}\,|\,(7) \&(12)\, hold.\};\,
subset (b)=\{g_{s1},g_{s2},g_{1d},g_{2d}\,|\,(8) \&(12)\, hold.\}
\end{eqnarray*}
\begin{eqnarray*}
subset (c)=\{g_{s1},g_{s2},g_{1d},g_{2d}\,|\,(9) \&(12)\, hold.\};\,
subset (d)=\{g_{s1},g_{s2},g_{1d},g_{2d}\,|\,(10) \&(12)\, hold.\}
\end{eqnarray*}
\begin{eqnarray*}
subset (e)=\{g_{s1},g_{s2},g_{1d},g_{2d}\,|\,(7) \&(13)\, hold.\};\,
subset (f)=\{g_{s1},g_{s2},g_{1d},g_{2d}\,|\,(8) \&(13)\, hold.\}
\end{eqnarray*}
\begin{eqnarray*}
subset (g)=\{g_{s1},g_{s2},g_{1d},g_{2d}\,|\,(9) \&(13)\, hold.\};\,
subset (h)=\{g_{s1},g_{s2},g_{1d},g_{2d}\,|\,(10) \&(13)\, hold.\}
\end{eqnarray*}

According to the above analysis, an effective hybrid scheme
combining the SRP and AFP modes is presented here.

\textbf{An Opportunistic Scheduling Scheme :} {\em In the diamond
relay network model, if the channel side information is obtained by
all the nodes at the beginning of each time block, it is possible to
find the thresholds properly for subspace partition and select an
effective cooperative mode with larger throughput.}

Note that the switch between the two cooperative models will not be
very frequent due to the block fading channel. It only occurs
between two time blocks if necessary.

\section{A New Processing Strategy With Buffers Employed by Relay Nodes}
\label{model3}

Numerical results in Fig. 5  indicated that SRP always performs much
better than AFP on average due to its full spatial reuse. This means
that only adopting a hybrid scheme simply can not bring an expected
throughput improvement.  Since the links associated to the relay is
 time-varying and block fading, some links in bad conditions will form a bottleneck, which greatly
 reduce the throughput. To overcome it, we shall put forward a new processing strategy in
which buffers are used at both relays to efficiently use the links
in good conditions, which is referred to opportunistic scheduling.

In fact, it is possible to find an effective opportunity scheduling
under some cases, such as that only one or two links are in very
good conditions while the others are all in bad conditions. Now we
consider the following two cases shown in Fig. 3.
\begin{enumerate}
\item At least one link from $N_s$ to the two relays are in pretty good
conditions while the two links from the relays to destination are in
rather bad conditions.
\item At least one link from the two relays to destination are in pretty good
conditions while the two links from source to the relays are in
rather bad ones.
\end{enumerate}

In Fig. 3,  symbol $G$ denotes the link  in pretty good conditions,
$B$ denotes it in rather bad ones and $X$ denotes it in arbitary
conditions  (i.e. pretty good, rather bad or average). In this two
cases, the links in rather bad conditions form the bottleneck of the
networks, especially for those schemes without buffers. Therefore,
the new opportunity scheduling scheme is expressed as follows.

\textbf{New Strategy With Buffers Employed }

 (1) Under the above tow
cases, we shall use these links with pretty good conditions to
transmit signals while the links with rather bad conditions will
keep silent and some received signals will be stored at its
corresponding buffers.

(2) For other cases, we still adopt the same policy as that in
hybrid scheme of combining SRP and AFP without buffers.

From the new opportunity scheduling scheme, one can find that a new
raising problem is how to determine each link condition being good
or bad? Furthermore, the evaluation of queuing delay caused by using
buffers is a new problem to be considered. In the sequel, we shall
deal with them.

\subsection{Two Criteria for Selecting  Threshold of Link Condition}

Since the signals transmitted from source to both relays in AFP mode
is the same and the destination needs to receive it coherently, no
buffers in AFP mode may loss some opportunities to adapt the link
time-varying conditions. For SRP mode, source will transmit
different signals to the two relays, the sequence number of the
received symbols at the destination may be different with its
original ones due to these link time varying capacities, which is
similar to the phenomenon in Internet. Furthermore, $C_{SR}$ is
mainly determined by the very bad links according to Theorem 3.2
when the conditions of the four links have much difference.
Therefore, buffers employed in SRP mode for the two cases listed
above can mitigate the impact of bottleneck links. In other cases,
e.g., the conditions of the links are almost the same, no matter
whether it is good or bad, the impact of bottleneck links is
relatively small for SRP mode.

To separate the pretty good and rather bad link states, we first
propose two thresholds of the link state levels, denoted as
$C_{upTH}$ and $C_{dwTH}$, respectively, based on the finite state
fading channel model. That is to say, the link is considered to be
in pretty good condition if its state level is above or equal to
$C_{upTH}$ and considered to be in rather bad condition if its state
level is below or equal to $C_{dwTH}$.

\begin{equation}
\textbf{Criterion 1:}\hspace{40mm} C_{upTH} > 2\cdot C_{dwTH}\hspace{30mm}
\end{equation}

Based on this criterion, we  shall prove that only under the two
cases discussed above, the throughput improvement can be obtained.
The detail of proof is given in Appendix (C). Here we shall give an
explanation for Criterion 1 in principle.

Consider two consecutive time blocks $T_1$ and $T_2$, $T_1=T_2=T$,
and $T_1$ is prior to $T_2$.  For the hybrid scheme without buffers,
the total amount of information transmitted in this two consecutive
time blocks, denoted as $THR_1$, is given by
\begin{eqnarray}
THR_1=T_1 \cdot \max\{C_{SR}(T_1),C_{AF}(T_1)\}+T_2 \cdot
\max\{C_{SR}(T_2),C_{AF}(T_2)\}
\end{eqnarray}
and for the new developed strategy, the total amount of information
transmitted in this two consecutive time blocks, denoted as $THR_2$,
is given by
\begin{eqnarray}
THR_2=T \cdot
\max\{\min\{C_{s1}(T_1),C_{1d}(T_2)\},\min\{C_{s2}(T_1),C_{2d}(T_2)\}\}
\end{eqnarray}

Note that  ``$\min\{C_{si}(T_1),C_{id}(T_2)\},\{i=1,2\}$" is to
guarantee that the buffer is not empty while ``$\max$" in Eqn. (16)
means utilizing the better route. In fact, in a causal system, one
can not obtain  $C_{id}(T_2)$ during the block $T_1$, thus Eqn.(16)
only presents an ideal case. In practice, we shall use the better
front-side link associated to the two relays when case (1) happens.
This is because the probabilities of different link states are equal
and the link states change independently from one block to next,
which means the two back-side links associated to the relays have
the same probability being in pretty good conditions when case (2)
happens. From statistic view of point, choosing the better
front-side link and sending massage as much as possible will obtain
a larger gain. In addition, if the buffer becomes empty as case (2)
happens, the relay will not transmit any massage. This is the case
just as that in a general G/G/1 system. Even though, the performance
will degrade little because the throughput of those non-buffer
strategies also become rather small when case (2) happens. Thus, we
can still use Eqn.(16) in the following analysis approximately.

Let us observe the case that in $T_1$ only the link from $N_s$ to
$N_1$ is pretty good and in $T_2$ only link from $N_1$ to $N_d$ is
pretty good while all the other three links are rather bad, which
means $C_{si}(T_1)\geq C_{upTH}$ and $C_{id}(T_2)\geq C_{upTH}$
hold. In this case, the throughput of the hybrid scheme without
buffers is very small due to the existence of bottleneck links. But
for the new developed strategy, since one buffer is employed by
$N_1$, the total throughput in the two time blocks becomes much
larger due to the efficient utilization of the link from $N_s$ to
$N_1$ in $T_1$ and the good link from $N_1$ to $N_d$ in $T_2$.

Associated with that SRP mode is better than AFP on average, it can
be concluded that if the inequality $THR_2> THR_1$ holds, then the
total amount of information transmitted in these two consecutive
time blocks with the buffer's help will be larger than that in the
original hybrid mode in a certain degree. For other cases except
cases $(1)$ and $(2)$, buffers will not be used and their
corresponding throughput parts keep the same.  Therefore, one can
see that the new developed strategy will improve the average
throughput. The proof of $THR_2> THR_1$ is given in Appendix (C).
Its main idea is that $C_{SR} \leq C_{dwTH}$ holds in these two
cases according to Theorem 3.1. If the condition $C_{upTH}
> 2\cdot C_{dwTH}$ in Criterion 1 is satisfied, the inequality $THR_2>
THR_1$ will hold and the improvement of throughput can be
guaranteed.

TABLE I shows the levels of 16-state partition of the fading channel
obtained by the corresponding maximum achievable rate with the
normalization ${P_c}/{\sigma^2}=1$ for different received SNR.
According to Criterion 1, $C_{dwTH}$ is selected from the left side
of $\ast$ for each link state level and $C_{upTH}$ is selected from
the right side of $\star$ for the corresponding state level. In
addition, the two sides are formed symmetrically because the
performance of original schemes is relatively good for the case that
link state levels fall into the interval between $\ast$ and $\star$
since the impact of the bottleneck only dominates when the link
states differ a lot.



\subsection{Delay Analysis}
In the previous example, the two time blocks, $T_1$ and $T_2$, are
assumed to be consecutive. In fact, they may not be adjacent to each
other according to the {\em$i.i.d$} link state model. That is, the
information in the buffer has to wait for transmission. On the other
hand, due to the links being in Rayleigh fading, the probability of
links in pretty good conditions is relatively low, resulting in a
larger delay for information transmission. In the new developed
strategy, the mean value of the delay for information transmission
is mainly determined by the thresholds $C_{upTH}$ and $C_{dwTH}$.
Based on the topological symmetry of the diamond relay networks, we
only analyze the performance of the subsystem, shown in Fig. 4.  In
Table II, some notations are firstly defined.

with the {\em$i.i.d$} link state model, $P_x$ and $P_y$ are given by
\begin{eqnarray}
P_x&=&\frac{N-U+1}{N}(\frac{U-1}{N}+\frac12 \frac{N-U+1}{N})(\frac{d}{N})^2=\frac{N^2-(U-1)^2}{2N^2}(\frac{d}{N})^2\nonumber\\
P_y&=&\frac{N-u+1}{N}(\frac{u-1}{N}+\frac12
\frac{N-u+1}{N})(\frac{D}{N})^2=\frac{N^2-(u-1)^2}{2N^2}(\frac{D}{N})^2
\nonumber
\end{eqnarray}
Note that the item $\frac{N-U+1}{N}(\frac12 \frac{N-U+1}{N})$ is the
probability that the condition of the link between $N_s$ and $R_1$
is better than that between $N_s$ and $sim R_2$ when they are both
pretty good. In the new developed strategy, only if the link state
falls into one of the two cases mentioned previously, the buffer
works. In other cases, the buffer will not work and the networks run
with the same procedure as the non-buffer scheme. That is,  no new
massage is put into the buffer though it is delivered to the
corresponding relay node. The massage  queuing in the buffer
previously will wait for the moment at which cases (1) or (2) happen
again. The details of the queuing model can be described as follows.
\begin{enumerate}
\item Arrival process: Since the source always
has information to deliver, the link state to the relay determines
the input process of the buffer, refer to the arrival process of
buffer, including the arrival interval and rate. When link state is
at level $k$, the arrival interval between the successive traffic
units is equal to $\varepsilon/C_k$. The symbol
$\varepsilon\in(0,1]$ is a parameter determined by the traffic types
of source (i.e. bit, byte or packet), which is not a key point and
for simplicity, let $\varepsilon=1$ represent a packet.
\item Service process: The service process
refers to the delivering process of the message stored in the buffer
of the relays to destination. Once a packet stored in the buffer is
transmitted successfully, it is served. Thus, the link state between
the relay and destination determines the service process, including
the service interval and rate.
\end{enumerate}

The arrival interval of traffic units needs to be considered for the
following three cases: when the buffer works, the condition of the
link to the relay is pretty good. When the buffer works, the
condition of the link to the relay is rather bad; and that buffer
does not work. The value of arrival interval and its corresponding
probability for each case is

\begin{equation*}\label{eq:1}
\left\{ \begin{aligned}
         1/C_i,\,\,&prob.\,\,equal\,\,to \,P_x/(N-U+1),\,\,i\in[U,N]\\
         1/C_j,\,\,&prob.\,\,equal\,\,to \,P_y/D,\,\,j\in[1,D]\\
         nT,\,\,&prob.\,\,equal\,\,to \,(1-P)^n P,\,\,n\in[1,+\infty),\,\,n\in\mathbb{\mathrm{Z}^+}
                          \end{aligned} \right.
                          \end{equation*}
and for the service time, similar results are given by
\begin{equation*}\label{eq:2}
\left\{ \begin{aligned}
         1/C_i,\,\,&prob.\,\,equal\,\,to\, P_y/(N-u+1),\,\,i\in[u,N]\\
         1/C_j,\,\,&prob.\,\,equal\,\,to \,P_x/d,\,\,j\in[1,d]\\
         nT,\,\,&prob.\,\,equal\,\,to \,(1-P)^n P,\,\,n\in[1,+\infty),\,\,n\in\mathbb{\mathrm{Z}^+}
                          \end{aligned} \right.
                          \end{equation*}

It is easy to see that the distributions of the arrival and service
interval do not obey the uniform or Poisson distribution,etc. They
are general. Therefore, one can solve it with the G/G/1 queuing
model. The mean values and variances are given in Theorem 4.1.

In fact, the stability of queueing in buffer is very important. To
solve this problem, the traffic intensity of buffer, denoted as
$\rho$, should be less than $1$. In the new strategy, the stability
of buffer can be guaranteed by selecting the thresholds, $C_{upTH}$
and $C_{dwTH}$, following criterion 2:
\begin{equation}
\textbf{Criterion 2:}\hspace{20mm}(i)\, U=u \,\,\,\,\, and\,\,\,\,\,
D>d \,\,;\,\,(ii)\,U>u \,\,\,\,\,and \,\,\,\,\, D=d \hspace{20mm}
\end{equation}
The proof is given in the Appendix (D). Furthermore, one can easily
infer that for the case ``$U\geq u \,\,\,\,\, and\,\,\,\,\, D>d$" or
``$U>u \,\,\,\,\, and\,\,\,\,\, D\geq d$", the buffer is also
stable. Besides, due to that the thresholds, $U$ and $u$ or $D$ and
$d$ are relatively close to each other and the source node has a
continuous traffic to deliver, the arrival rate of the buffer is
then close to but smaller than the service rate. That is, the
traffic intensity of the buffer, the ratio of the arrival rate to
the service rate, approach to $1$. Consequently, the upper bound of
the average delay in Marshall's theory become more effective.

\textbf{Theorem $4.1$:} For the new developed strategy with buffers
at both relay nodes, we have
\begin{eqnarray}
E(a)=E(t_U)P_x+E(t_D)P_y+((1-P)/P)T,\,
E(b)=E(t_u)P_y+E(t_d)P_x+((1-P)/P)T
\end{eqnarray}
\begin{eqnarray}
\sigma_a^2&=&(E(t_U^2)+E^2(a)-2E(a)E(t_U))P_x +(E(t_D^2)+E^2(a)-2E(a)E(t_D))P_y \nonumber \\
&&+(1-P)E^2(a)+(\frac{1-P}{P})2T\cdot E(a)
+(\frac{(1-P)(2-P)}{P^2})T^2
\end{eqnarray}
\begin{eqnarray}
\sigma_b^2&=&(E(t_u^2)+E^2(b)-2E(b)E(t_u))P_y +(E(t_d^2)+E^2(b)-2E(b)E(t_d))^2P_x \nonumber \\
&&+(1-P)E^2(b)+(\frac{1-P}{P})2T\cdot E(b)
+(\frac{(1-P)(2-P)}{P^2})T^2
\end{eqnarray}
\begin{eqnarray}
&\overline{W}&\leq \frac{\sigma_a^2+\sigma_b^2}{2(E(a)-E(b))}
\end{eqnarray}
where $E(t_U)$ , $E(t_D)$ , $E(t_u)$ and $E(t_d)$ are given by,
respectively,
\begin{equation*}
E(t_U)=\DF{\sum_{i=U}^N 1/C_i}{N-U+1} \,,\, E(t_D)=\DF{\sum_{i=1}^D
1/C_i}{D}\,,\,E(t_u)=\DF{\sum_{i=u}^N 1/C_i}{N-u+1} \,,\,
E(t_d)=\DF{\sum_{i=1}^d 1/C_i}{d}
\end{equation*}
The proof is given in the Appendix (E). $Theorem\,\,4.1$, presented
the upper bound of the average delay, $\overline{W}$, which is a
function of the thresholds $U$ , $D$, $u$ and $d$. Consequently, it
provides an theoretical way to consider the good tradeoff between
the average delay and the network throughput.

\subsection{Tradeoff Between the Throughput and the Delay}
Follow the above discussions in Subsections IV. A and IV. B, the new
developed strategy can really improve the throughput by efficiently
utilizing the pretty good links at a cost of queuing delay. Using
the two criteria, for the fixed $D$ and $d$, if a relatively lower
$u$ and higher $U$ are selected, the throughput improvement is less
than that with selection of a relatively higher $u$ and lower $U$.
This is because that in the former case the value of $(P_x P_y)^2
(THR_2-THR_1)$ is less than the one in the latter case (see Appendix
(F)). On the other hand, according to $Theorem\,\,4.1$, the former
case has a shorter delay due to its higher service rate.  A similar
result can be observed for selecting $D$ and $d$ under the condition
that $U$ and $u$ are fixed. Simulations in Section \ref{model4} will
also confirm this phenomenon.

To achieve a good tradeoff between the throughput and the delay, we
shall present an algorithm to select the thresholds, $C_{upTH}$ and
$C_{dwTH}$, which is summarized as follows.

\textbf{Enumerative Algorithm :}{\em Due to that the nodes are able
to know the partition of link states in advance and obtain the
average received SNR at the beginning of each time block $T$. In
other words, they already have the side information in Table I.}
\begin{enumerate}
\item Step 1: Enumerate the combinations of $C_{upTH}$ and
$C_{dwTH}$ according to both of the criteria.
\item Step 2: Estimate the average delay for each case via
$Theorem\,\,4.1$ and find out all the possible combinations of
$C_{upTH}$ and $C_{dwTH}$ whose average delay is shorter than the
requirement by the service traffic. Let us denote the available set
as $\Gamma$.
\item Step 3: For all the possible pair of $C_{upTH}$ and $C_{dwTH}$
belonging to $\Gamma$, to achieve larger throughput the nodes select
$U$ as low as possible and $u$ as high as possible or select $D$ as
low as possible and $d$ as high as possible.
\end{enumerate}

\section{Numerical Results}
\label{model4} In this section, simulation is used to demonstrate
our theoretical results. In the simulation part, the stream traffic
model is employed and the source node is assumed to have massage to
deliver always. The average received SNR of these two relays and the
destination varies from 0dB to 10dB. The simulation period  consists
of $10^5\sim10^6$ time blocks, denoted by $M$, where each time block
$T$ is equal to $1ms$.  For the new developed strategy, the total
amount of information received successfully by the destination is
$R_{suc}$. Then, the average throughput of the new developed
strategy is evaluated by $R_{suc}/M$. Since there are several
selections of $C_{upTH}$ and $C_{dwTH}$ for each average received
SNR, we estimate the average throughput for each case under the same
average received SNR. For the original schemes, similar procedure
are done.


Fig. 5 shows the maximum achievable rates of SRP, AFP, the hybrid
scheme and the new developed strategy with buffers for each average
received SNR. Both the theoretical and simulation results indicate
that if all the links have the same average SNR, SRP mode is always
better than AFP mode. Consequently, the hybrid scheme simply
combining both of them is almost equivalent to that only adopting
SRP mode. Compared with the original schemes, the new developed
strategy with buffers really improve the average throughput, e.g.
 there is an approximate increment of $0.071$ unit/s
at the average received SNR of 4dB. That is about $11\%$
improvement. In addition, the improvement rate will decrease as the
received SNR increases further. It is because that when the average
capacity of each link increases under a higher SNR, the degree of
the bottleneck link influence becomes smaller.


Fig. 6 illustrates the average delay under different received SNR,
evaluated in terms of the number of $T$. It is shown that for a
fixed average received SNR, the average delay increases as the
traffic intensity becomes larger and for a fixed traffic intensity,
the average delay decreases as the average received SNR increases.
For instance, when traffic intensity is $0.98$, the delay in the
average received SNR of 2dB is $69T$, while the delay in the average
received SNR of 10dB, it is only $19T$. Also it can be seen that the
simulated curve becomes much closer to the theoretical one as $\rho$
approaches to 1, which demonstrated that the upper bound derived in
theory is effective.

Fig. 7 shows the tradeoff between the improvement of network
throughput and the average delay, which is consistent with our
theoretical predication. Some simulation results are also listed in
Table III where the average received SNR is equal to $6dB$. It is
shown that if the required average delay by the traffic is no more
than 20 time blocks, one can only obtain a throughput increment of
0.0311bit/s when the thresholds are selected as $D=3$, $d=2$, $U=16$
and $u=15$. That is about $4\%$ improvement compared to the original
value in Table I. A larger improvement can be obtained if the
requirement of average delay becomes loose. One can see that when
nodes select a lower $U$ and a higher $u$ or select a lower $D$ and
a higher $d$, the throughput improvement become larger while the
delay becomes larger simultaneously. In addition, if the average
delay approaches to infinity in the case that $D=2$, $d=2$, $U=15$
and $u=15$, since it does not match the stable conditions, Criterion
2.


\section{Conclusions}
\label{model5}

In this paper, the diamond relay mode was studied. We compared the
throughput performance of the two classical cooperative modes, SRP
(Spatial Reused Pattern) and AFP (Amplify Forwarding Pattern) under
wireless scenarios with assumption that the channel side information
is known to all the nodes. We analyzed the possibility to improve
the throughput by employing buffers at relays and proposed an new
opportunity scheduling scheme. In order to improve the network
throughput while guaranteing the stable running of the whole
network, we established two criteria on the selection of SNR
thresholds and one adjustment algorithm on the tradeoff between the
throughput improvement and the queueing delay. Simulation results
confirmed the effectiveness of our theoretical analysis and our new
developed opportunity scheduling method.

\section*{Acknowledgment}
It was supported by NSFC/RGC Joint Research Scheme No.60831160524
and the open research fund of National Mobile Communications
Research Laboratory, Southeast University, China.

\section*{Appendix (A)}

\textbf{Proof of Theorem $3.2$:} The original expression can be
transformed into the equivalent form:
\begin{eqnarray*}
C_{AF}&=&\max_{\alpha,\beta}\{\frac12 \cdot \frac12\log
(1+\frac{(||g_{s1}||^2+ ||g_{s2}||^2)}{1+\DF{1}{\alpha^2+\beta^2}}
\frac{P_c}{\sigma^2}) \}
\end{eqnarray*}
\begin{eqnarray*}
s.t.\,\,\alpha^2\leq (1+\frac{||g_{s1}||P_c}{\sigma^2})/
(\frac{||g_{1d}||P_c}{\sigma^2}),\,\beta^2\leq
(1+\frac{||g_{s2}||P_c}{\sigma^2})/(\frac{||g_{2d}||P_c}{\sigma^2}),\,\alpha^2
/ \beta^2 = ||g_{s1}||^2 / ||g_{s2}||^2.
\end{eqnarray*}
That means $C_{AF}$ achieves its maximum when $\alpha^2+\beta^2$
reaches its maximum under the three constraints above. With the
three constraints, one can find that
\begin{eqnarray}
\alpha^2+\beta^2=(1+\frac{||g_{s1}||^2}{||g_{s2}||^2})\beta^2 &\leq&\min\{\frac{||g_{2d}||^2(||g_{s1}||^2+ ||g_{s2}||^2)}{||g_{s2}||^2(||g_{s2}||^2+\frac{\sigma^2}{P_c})}, \frac{||g_{1d}||^2(||g_{s1}||^2+ ||g_{s2}||^2)}{||g_{s1}||^2(||g_{s1}||^2+\frac{\sigma^2}{P_c})}\} \nonumber\\
&=&\min\{\frac{G_{2d}(G_{s1}+G_{s2})}{G_{s2}(G_{s2}+1)},\frac{G_{1d}(G_{s1}+G_{s2})}{G_{s1}(G_{s1}+1)}\}
\end{eqnarray}
Thus, the expression of $C_{AF}$ in Eqn.(22) is obtained.
$\hspace{1mm} \square$

\section*{Appendix (B)}

\begin{eqnarray}
C_{AF}&=&\max_{\alpha,\beta}\{\frac12 \cdot \frac12\log
(1+\frac{(\alpha ||g_{s1}||+\beta ||g_{s2}||)^2}{\alpha^2+\beta^2+1}
\frac{P_c}{\sigma^2}) \} \nonumber \\
&s.t.&\alpha^2(1+\frac{||g_{s1}||P_c}{\sigma^2})\leq
\frac{||g_{1d}||P_c}{\sigma^2}\,,\,\beta^2(1+\frac{||g_{s2}||P_c}{\sigma^2})\leq
\frac{||g_{2d}||P_c}{\sigma^2}.\nonumber
\end{eqnarray}
The original problem above is equivalent to the following one,
\begin{equation}
(x_{opt},y_{opt})=\mathop {\arg }\limits_{x,y} \max F,
\,\,s.t.0<x\leq C;0<y\leq D;A,B,C,D>0
\end{equation}
where
\begin{eqnarray}
F=\frac{(Ax+By)^2}{x^2+y^2+1},x=\alpha,y=\beta,A=||g_{s1}||,B=||g_{s2}||
C= \sqrt{\frac{||g_{1d}||^2}{\frac{\sigma^2}{P_c}+||g_{01}||^2}},D=
\sqrt{\frac{||g_{2d}||^2}{\frac{\sigma^2}{P_c}+||g_{02}||^2}}
\nonumber
\end{eqnarray}
To derive the maximum value of $F$, we first have
\begin{eqnarray}
\frac{\partial F}{\partial
x}=\frac{2(Ax+By)}{x^2+y^2+1}[A-\frac{x(Ax+By)}{x^2+y^2+1}]\,,\,\frac{\partial
F}{\partial
y}=\frac{2(Ax+By)}{x^2+y^2+1}[B-\frac{y(Ax+By)}{x^2+y^2+1}]
\end{eqnarray}
then we solve the following equations,
\begin{eqnarray}
\frac{\partial F}{\partial x}=0 \Longrightarrow A+Ay^2=Bxy \Longrightarrow x=\frac{A}{B}(\frac{1+y^2}{y}) \leq \frac{2A}{B}\\
\frac{\partial F}{\partial y}=0 \Longrightarrow B+Bx^2=Axy
\Longrightarrow y=\frac{B}{A}(\frac{1+x^2}{x}) \leq \frac{2B}{A}
\end{eqnarray}
Rewriting Eqn. (23), it is easily found that
\begin{eqnarray}
0<x<\frac{A}{B}(\frac{1+y^2}{y}) \Longrightarrow \frac{\partial
F}{\partial x}>0 \,,\, x>\frac{A}{B}(\frac{1+y^2}{y})
\Longrightarrow \frac{\partial F}{\partial x}<0 \nonumber
\end{eqnarray}
This indicates that $F$ is a monotone increasing function of $x$ if
$x\in(0,\frac{A}{B}(\frac{1+y^2}{y}))$ and a monotone decreasing
function of $x$ if $x > \frac{A}{B}(\frac{1+y^2}{y})$. Similar
results can be derived for $\frac{\partial F}{\partial y}$.

In addition, we find that the two equations $\frac{\partial
F}{\partial x}=0$ and $\frac{\partial F}{\partial y}=0$ can not hold
simultaneously. Otherwise, $B^2=-A^2(1+\frac{1}{y^2})$, which
resulting in  contradiction.  In fact, if $\frac{\partial
F}{\partial x}=0$ holds, then $\frac{\partial F}{\partial y}>0$ will
be guaranteed.

According to the expression of $\frac{\partial F}{\partial y}$, we
only need to prove $\frac{B}{A}(1+x^2)-xy>0$ holds. Since
\begin{eqnarray}
 \frac{\partial F}{\partial x}=0 \Longrightarrow
x=\frac{A}{B}(\frac{1+y^2}{y}),\nonumber
\end{eqnarray}
we have
\begin{eqnarray}
\frac{B}{A}(1+x^2)-xy=\frac{B}{A}+\frac{A}{B}(\frac{1+y^2}{y^2})>0
\nonumber
\end{eqnarray}
Similar result can be obtained for that case if $\frac{\partial
F}{\partial y}=0$ holds, then $\frac{\partial F}{\partial x}>0$ will
be true.

In addition, according to Eqn.(25) and (26), if $2A>BC$ holds, then
$x\mid_{\frac{\partial F}{\partial x}=0}>C$ is true and if
$2B>AD$ holds, then $y\mid_{\frac{\partial F}{\partial y}=0}>D$ is
true. Now we can summarize different cases:

(1) For the case that $2A\leq BC$:

 If $\frac{A}{B}(\frac{1+D^2}{D})\leq C$ , then
$x_{opt}=\frac{A}{B}(\frac{1+D^2}{D})$ and $y_{opt}=D$. Otherwise,
we divide it into two subcases: a) If $BCD\geq A+AD^2$ holds, we
have $x_{opt}=C$ and $y_{opt}=\mathop {\max }\limits_{ \le C} \{
\frac{{BC + \sqrt {B^2 C^2  - 4A^2 } }}{{2A}},\frac{{BC - \sqrt {B^2
C^2  - 4A^2 } }}{{2A}}\}$. b) If $BCD<A+AD^2$ holds, we have
$x_{opt}=C$ and $y_{opt}=D$. where ``$\mathop {\max }\limits_{ \le
z} \{x,y\}$" denotes the larger one between $x$ and $y$ with the
constraint that both of them are less or equal to $z$. If anyone of
them is above $z$, its value is defined as $0$.

(2) For the case that  $2B\leq AD$ :

 If
$\frac{B}{A}(\frac{1+C^2}{C})\leq D$, then we have $x_{opt}=C$ and
$y_{opt}=\frac{B}{A}(\frac{1+C^2}{C})$. Otherwise, we also divide it
into two subcases. a) If $ACD\geq B+BC^2$, then we have $x_{opt}=
\mathop {\max }\limits_{ \le D} \{ \frac{{AD + \sqrt {A^2 D^2  -
4B^2 } }}{{2B}},\frac{{AD - \sqrt {A^2 D^2  - 4B^2 } }}{{2B}}\}$ and
$y_{opt}=D$. b) If $ACD<B+BC^2$, then we have $x_{opt}=C$ and
$y_{opt}=D$.

(3) For other cases: $F$ achieves the maximum value if $x_{opt}=C$
and $y_{opt}=D$. $\hspace{10mm} \square$

\section*{Appendix (C)}

Without loss of generality, let us consider the proof of case (a)
shown in Fig. 3. ( i.e. At least a link from source to the relays is
pretty good and the two links from the relays to destination are
rather bad.) It is easily to check that the condition (9) in Theorem
3.1 is satisfied for case (a).  Thus, the maximum achievable
transmitting rate can be given by
\begin{equation}
C_{SR}=\frac{C_{1d}(C_{s1}+C_{2d})}{C_{s1}+C_{1d}}
\end{equation}
where $C_{s1}\geq C_{upTH}$ , $C_{1d}\leq C_{dwTH}$ and $C_{2d}\leq
C_{dwTH}$ holds in case (a).

In addition, $C_{SR}$ is a monotone increasing function of both
$C_{1d}$ and $C_{2d}$. Thus, we have
\begin{equation}
C_{SR}<\frac{C_{dwTH}(C_{s1}+C_{dwTH})}{C_{s1}+C_{dwTH}}=C_{dwTH}
\end{equation}

Likewise, similar proofs can be given for another three cases in
Fig. 3. Finally, based on Criterion 1, we conclude that
$C_{SR}<C_{dwTH}$ holds only in the cases (1) and (2) presented in
Section IV.  $\hspace{10mm} \square$

\section*{Appendix (D)}

Firstly we deduce the average arrival rate and service rate for the
system shown in Fig. 4, denoted as $\lambda$ and $\mu$,
respectively. According to the new developed strategy, only cases
(1) or (2) happens, the buffers start to work. Therefore,  the
average transmitting rate when the link states are in pretty good
and rather bad conditions can be derived as follow
\begin{eqnarray}
\overline{C_H^U}=\frac{\sum_{i=U}^N
C_i}{N-U+1},\overline{C_H^u}=\frac{\sum_{i=u}^N
C_i}{N-u+1},\overline{C_L^D}=\frac{\sum_{i=1}^D
C_i}{D},\overline{C_L^d}=\frac{\sum_{i=1}^d C_i}{d} \nonumber
\end{eqnarray}
Associated with the probabilities that the buffers is in working
status, $P_x$ and $P_y$, we have
\begin{eqnarray}
\lambda=\overline{C_H^U}P_x+\overline{C_L^D}P_y =\overline{C_H^U}
\frac{N^2-(U-1)^2}{2N^2}(\frac{d}{N})^2+\overline{C_L^D}
\frac{N^2-(u-1)^2}{2N^2}(\frac{D}{N})^2
\end{eqnarray}
\begin{eqnarray}
\mu=\overline{C_H^u}P_y+\overline{C_L^d}P_x =\overline{C_H^u}
\frac{N^2-(u-1)^2}{2N^2}(\frac{D}{N})^2+\overline{C_L^d}
\frac{N^2-(U-1)^2}{2N^2}(\frac{d}{N})^2
\end{eqnarray}

Next we consider the two conditions expressed in Criterion 2,
respectively.

(i) If $U>u$ and $D=d$,
$\overline{C_L^D}=\overline{C_L^d}=\overline{C_L}$ holds, then we
have
\begin{eqnarray}
\lambda-\mu&=&(\frac{D}{N})^2 [\overline{C_H^U}\frac{N^2-(U-1)^2}{2N^2}-\overline{C_H^u} \frac{N^2-(u-1)^2}{2N^2}+\overline{C_L}(\frac{N^2-(u-1)^2}{2N^2}-\frac{N^2-(U-1)^2}{2N^2})] \nonumber\\
&<&(\frac{D}{N})^2 [\overline{C_H^U} \frac{N^2-(U-1)^2}{2N^2}-\overline{C_H^u} \frac{N^2-(u-1)^2}{2N^2}+(\frac{\overline{C_H^U}+\overline{C_H^u}}{2})(\frac{N^2-(u-1)^2}{2N^2}\nonumber\\
&&-\frac{N^2-(U-1)^2}{2N^2})]=2(\frac{D}{N})^2(\overline{C_H^U}-\overline{C_H^u})(\frac{N^2-(U-1)^2}{2N^2}-\frac{N^2-(u-1)^2}{2N^2})
<0 \nonumber
\end{eqnarray}
The first inequality above holds because $\overline{C_L}\leq
C_{dwTH} < \frac12 C_{upTH}\leq \frac12 \overline{C_H^u}< \frac12
\overline{C_H^U}$ is true. Associated with
$\overline{C_H^U}-\overline{C_H^u} >0$ and
$\frac{N^2-(U-1)^2}{2N^2}-\frac{N^2-(u-1)^2}{2N^2} <0$, one can see
that  $\lambda<\mu$. Thus, the corresponding traffic intensity is
less than 1, indicating that the queueing is stable.

(ii) If $U=u$ and $D>d$,
$\overline{C_H^U}=\overline{C_H^u}=\overline{C_H}$ holds, then we
have
\begin{eqnarray}
&\lambda-\mu&=\frac{N^2-(U-1)^2}{2N^2} \{\overline{C_H}[(\frac{d}{N})^2-(\frac{D}{N})^2]+\overline{C_L^D} (\frac{D}{N})^2-\overline{C_L^d} (\frac{d}{N})^2\} \nonumber\\
&&<\frac{N^2-(U-1)^2}{2N^2} \{(\overline{C_L^D}+\overline{C_L^d})[(\frac{d}{N})^2-(\frac{D}{N})^2]+\overline{C_L^D} (\frac{D}{N})^2-\overline{C_L^d} (\frac{d}{N})^2\} \nonumber\\
&&=\frac{N^2-(U-1)^2}{2N^2}
[\overline{C_L^D}(\frac{d}{N})^2)-\overline{C_L^d}(\frac{D}{N})^2)]
<0 \nonumber
\end{eqnarray}
Similarly, the first inequality above holds because
$\overline{C_L^d}< \overline{C_L^D} \leq C_{dwTH} < \frac12
C_{upTH}\leq \frac12 \overline{C_H}$ is true. Check all the possible
combinations in Table I according to Criterion 1 and $U=u,D>d$ for
each $SNR$ case, one can see that
$\overline{C_L^D}(\frac{d}{N})^2)-\overline{C_L^d}(\frac{D}{N})^2)
<0$ is true.

Likewise, we can also verify the case of the 8-level and 32-level
partition and  make sure the conclusion holds based on the two
Criteria. The explanation in principle is that the impact of the
variation of the terms ``$D^2$" and ``$d^2$" to the plus-minus of
the inequality is much more greatly than that exerted by the gap
between the $C_i$'s value in $[C_1,C_{dwTH}]$ for each case with
equal partition. Thus, $\lambda<\mu$ holds, which guarantees the
corresponding traffic intensity is less than 1 and the queue is also
stable. $\hspace{5mm} \square$

\section*{Appendix (E)}

\textbf{Proof of Theorem $4.1$:} Here just give the proof of the
mean and variance of arrival interval $a$, respectively, denoted as
$m(a)$ and $\sigma_a^2$. Similar proof for the service time can follow this one.\\
(1) When the buffer works and the link state is higher or equal to
state level $U$, we have
\begin{eqnarray}
E(t_{U_1})=\sum_{i=U}^N \frac{1}{C_i} \cdot
P\{state\,\,level=i\}=\sum_{i=U}^N \frac{1}{C_i} \cdot
\frac{1}{N}\frac{N+U-1}{2N}(\frac{d}{N})^2 =P_x \cdot\overline{t_U}
\end{eqnarray}
(2) When the buffer works and the link state is lower or equal to
the state level $D$, we have
\begin{eqnarray}
E(t_{U_2})=\sum_{j=1}^D \frac{1}{C_j} \cdot
P\{state\,\,level=j\}=\sum_{j=1}^D \frac{1}{C_j} \cdot \frac{1}{N}
\frac{N^2-(u-1)^2}{2N^2} \frac{D}{N}=P_y \cdot\overline{t_D}
\end{eqnarray}
(3) When the buffer does not work in the consecutive $n$ time
blocks, we have
\begin{eqnarray}
E(t_{U_3})=\sum_{n=1}^\infty n T \cdot P\{no\,\,arrival\,\, in\,\,
continuous\,\, n T \}=\sum_{n=1}^\infty  n T \cdot(1-P)^n P
=(\frac{1-P}{P})T
\end{eqnarray}
With the results above, we get $E(a)=E(t_{U_1})+E(t_{U_2})+E(t_{U_3})$.

The proof for the variance of the arrival interval is similar. For
example, When the buffer works and the link state is higher or equal
to state level $U$, we have
\begin{eqnarray}
\sigma_{t_{U_1}}^2&=&\sum_{i=U}^N (\frac{1}{C_i}-E(a))^2
\cdot P\{state\,\,level=i\}=(\sum_{i=U}^N (\frac{1}{C_i})^2+(N-U+1)E^2(a)-2E(a)\cdot \nonumber \\
&&\sum_{i=U}^N \frac{1}{C_i}) \cdot
\frac{1}{N}\frac{N+U-1}{2N}(\frac{d}{N})^2=P_x \cdot
(E(t_U^2)+E^2(a)-2E(a)E(t_U))
\end{eqnarray}
When the buffer works and the link state is lower or equal to state
level $D$, the proof of $\sigma_{t_{U_2}}^2$ is similar and we omit
it here. When the buffer does not work in the consecutive $n$ time
blocks, we have
\begin{eqnarray}
\sigma_{t_{U_3}}^2&=&\sum_{n=1}^\infty [(n T-E(a))^2 \cdot P\{no\,\,arrival\,\, in\,\, continuous\,\, n T \}] \\
&=&\sum_{n=1}^\infty[(n T-E(a))^2 \cdot(1-P)^n
P]=\sum_{n=1}^\infty[(n^2T^2+E^2(a)-2nT\cdot E(a))\cdot(1-P)^n
P]\nonumber
\end{eqnarray}
By using the known results, for $q\in[0,1)$, $\sum_{n=1}^\infty n^2
q^n=\frac{q(1+q)}{(1-q)^3}$,

and

\begin{eqnarray}
\sum_{n=1}^m n^2 q^n&=&\frac{1}{(q-1)^3}\cdot[q(-1-q+q^m+2mq^m+m^2q^m \nonumber \\
&&+q^{1+m}-2mq^{1+m}-2m^2 q^{1+m}+m^2q^{2+m})],
\end{eqnarray}
we have
\begin{equation}
\sigma_{t_{U_3}}^2=(1-P)E^2(a)+(\frac{1-P}{P})2T\cdot E(a)
+(\frac{(1-P)(2-P)}{P^2})T^2
\end{equation}
Finally, we get $\sigma_a^2=\sigma_{t_{U_1}}^2+\sigma_{t_{U_2}}^2+\sigma_{t_{U_3}}^2$.$\hspace{10mm} \square$

\section*{Appendix (F)}
Consider the average throughput improvement in the two time blocks
$T_1$ and $T_2$ mentioned in our discussion. It can be denoted as
$P\{buffer,\,works\}\cdot(THR_2-THR_1)$. In addition, according to
the expression of $THR_2$, we need to find out the transmitting rate
(i.e.the explicit value of $min\{C_{si}(T_1),C_{id}(T_2)\}$ ) in all
the possible cases and their corresponding probabilities.

For the fixed $D$ and $d$, if we choose $U\in[C_{upTH}+1,N]$, then
$u\in[C_{upTH},U]$ subjected to Criterion 2. The probability that
the buffer works, $P_{new}$, in the two time blocks $T_1$ and $T_2$
is
\begin{equation}
P_{new}=P_xP_y=\frac{N^2-(U-1)^2}{2N^2}(\frac{d}{N})^2\cdot
\frac{N^2-(u-1)^2}{2N^2}(\frac{D}{N})^2
\end{equation}

In addition, if $C_{si}(T_1)>C_{id}(T_2)$ holds, the average
transmitting rate, denoted as $A$, is
\begin{equation}
A=\sum_{i=u}^{U-1} \frac{1}{N-u+1} C_i + \sum_{i=U}^{N-1}
\frac{N-i}{(N-U+1)(N-u+1)} C_i
\end{equation}
and if $C_{si}(T_1)\leq C_{id}(T_2)$ holds, the average transmitting
rate, denoted as $B$, is given by
\begin{equation}
B=\sum_{j=U}^{N} \frac{N-j+1}{(N-U+1)(N-u+1)} C_j
\end{equation}

Summarizing the results above, we have $THR_2=(A+B)\cdot T$. That
is,
\begin{equation}
THR_2=[\frac{1}{N-u+1} \sum_{k=u}^{U-1} C_k+(\sum_{k=U}^{N-1}
(2N-2k+1)C_k+C_N)\frac{1}{(N-U+1)(N-u+1)}]\cdot T
\end{equation}

Let $\Psi_{(U,u)}$ represent the throughput improvement when $U$ and
$u$ are selected. Then
\begin{eqnarray}
\Psi_{(U,u)}&=&P_{new} (THR_2-THR_1)=(\frac{D}{N}\frac{d}{N})^2 \frac{N^2-(U-1)^2}{2N^2}\frac{N^2-(u-1)^2}{2N^2} \cdot \nonumber\\
&&[\frac{1}{N-u+1} \sum_{k=u}^{U-1} C_k+\frac{\sum_{k=U}^{N-1}
(2N-2k+1)C_k+C_N}{(N-U+1)(N-u+1)}-{THR_1}/{T}]\cdot T
\end{eqnarray}

Next we keep $u$ as a invariable. For the case that
$U^\prime=U+m\,(1\leq m\leq N-U)$, and we have
\begin{eqnarray}
\Psi_{(U+m,u)}&=&P_{new} (THR_2-THR_1)=(\frac{D}{N}\frac{d}{N})^2 \frac{N^2-(U+m-1)^2}{2N^2}\frac{N^2-(u-1)^2}{2N^2} \cdot \nonumber \\
&&\{\frac{1}{N-u+1} \sum_{k=u}^{U+m-1} C_k +
\frac{\sum_{k=U+m}^{N-1}(2N-2k +1) C_k
+C_N}{(N-U-m+1)(N-u+1)}-{THR_1}/{T}\} \cdot T
\end{eqnarray}

In order to prove that $\Psi_{(U,u)}>\Psi_{(U+m,u)}$, we use the
following way.

Based on $C_u<C_U<C_{U+m-1}\leq C_{N-1}<C_N$, in the expression of
$\Psi_{(U,u)}$, we replace the former items $C_k\,(k=u,...,U-1)$
with $C_u$ and the latter items $C_k\,(k=U,...,N-1)$ with $C_U$,
then we obtain the infimum of $\Psi_{(U,u)}$ as follows
\begin{eqnarray}
\Psi_{(U,u)}^{inf}&=&\delta [N^2-(U-1)^2]\{\frac{U-u}{N-u+1} C_u + \nonumber \\
&&\frac{1}{(N-U+1)(N-u+1)}[(N-U)(N-U+2)C_U+C_N]-{THR_1}/{T}\}.
\end{eqnarray}

In the expression of $\Psi_{(U+m,u)}$, we replace the former items
$C_k\,(k=u,...,U+m-1)$ with $C_{U+m-1}$ and the latter items
$C_k\,(k=U+m,...,N-1)$ with $C_{N-1}$, then we get supremum of
$\Psi_{(U+m,u)}$ as follows
\begin{eqnarray}
\Psi_{(U+m,u)}^{sup}&=&\delta\cdot
[N^2-(U+m-1)^2][\frac{U+m-u}{N-u+1} C_{U+m-1}+\frac{1}{(N-U-m+1)(N-u+1)}C_N\nonumber\\
&&+\frac{(N-U-m)(N-U-m+2)}{(N-U-m+1)(N-u+1)}
C_{N-1}-{THR_1}/{T}]
\end{eqnarray}
where $\delta=T(\frac{1}{2N^2}\frac{D}{N}\frac{d}{N})^2
[N^2-(u+1)^2]$.

Now by scaling $\Psi_{(U,u)}^{inf}$ and $\Psi_{(U+m,u)}^{sup}$
further according to $C_u<C_U<C_{U+m-1}\leq C_{N-1}<C_N$, we have
\begin{eqnarray}
\Psi_{(U,u)}^{inf}&>&\delta\cdot [N^2-(U-1)^2]\{[\frac{(N-U)(N-U+2)}{(N-U+1)(N-u+1)} \nonumber\\
&&+\frac{U-u}{N-u+1}+\frac{1}{N-U+1}\cdot\frac{1}{N-u+1}]C_u-{THR_1}/{T}\} \\
&=&\delta\cdot [N^2-(U-1)^2](C_u-{THR_1}/{T})=\Psi_1 \nonumber
\end{eqnarray}
\begin{eqnarray}
\Psi_{(U+m,u)}^{sup}&<&\delta\cdot
[N^2-(U+m-1)^2][(\frac{U+m-u}{N-u+1}+\frac{(N-U-m)(N-U-m+2)}{(N-U-m+1)(N-u+1)}\nonumber\\
&&+\frac{1}{(N-U-m+1)(N-u+1)}) C_N-{THR_1}/{T}] \nonumber\\
&=&\delta\cdot [N^2-(U+m-1)^2](C_N-{THR_1}/{T})=\Psi_2
\end{eqnarray}

From the deductions above, the proof of
$\Psi_{(U,u)}>\Psi_{(U+m,u)}$ is equivalent to the proof of
$\Psi_1>\Psi_2$.

 Using the result in  Appendix (D), we have
$0<{THR_1}/{T}\leq 2C_{dwTH}$. From Table 1, we find that
$(C_u-{THR_1}/{T})>\alpha (C_N-{THR_1}/{T})$ always holds for all
the cases of average SNR if $\alpha= \frac23$. Furthermore,
\begin{eqnarray}
\Psi_1-\Psi_2&>& \{\alpha[N^2-(U-1)^2]-[N^2-(U+m-1)^2]\} \cdot
\delta\cdot (C_N-{THR_1}/{T})
\end{eqnarray}

This indicates that to prove $\Psi_1>\Psi_2$, we only need to
analyze the expression $\alpha[N^2-(U-1)^2]-[N^2-(U+m-1)^2]$. In
fact, if the following inequality holds,
\begin{equation}
m\geq \sqrt{\alpha(U-1)^2+(1-\alpha)N^2}+1-U
\end{equation}
the inequality $\alpha[N^2-(U-1)^2]-[N^2-(U+m-1)^2]\geq 0$ will be
true.

According to Table 1, we obtain that the maximum value of
$\sqrt{\alpha(U-1)^2+(1-\alpha)N^2}+1-U$ is equal to 0.342 when
$U=14$ for each case of SNR. Thus, inequality (49) is guaranteed for
$m\in[1,N-U]$. Likewise, similar proof can be given in the case that
$U$ is first selected and $u$ varies. It can be proved that when $u$
decreases, the improvement of the throughput will become smaller.

Thus, decreasing $U$ and increasing $u$ subjected to our established
criteria will bring a larger throughput.

In addition, one can see that the procedure of proof is related to
the values in Table 1, which means it is influenced by the levels of
link state partition. In fact, we can also verify the case of the
8-level and 32-level partition and make sure the conclusion also
holds.

\begin{figure}[h]
\centering
\includegraphics[width = 5in]{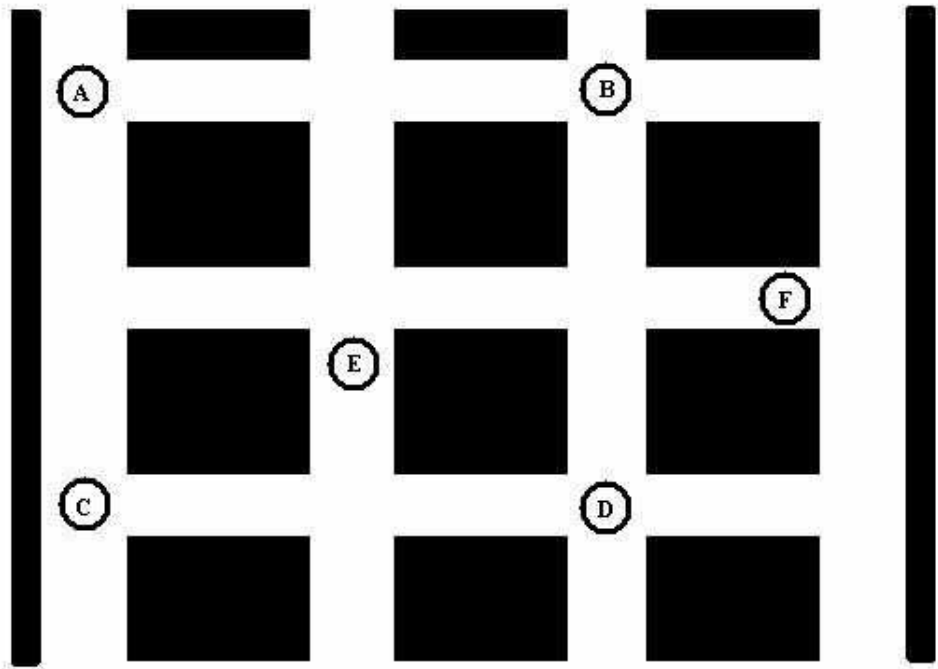}
\centering \caption{ Application Scenario Example.} \label{fig_1}
\end{figure}

\begin{figure}[h]
\centering
\includegraphics[width = 5in]{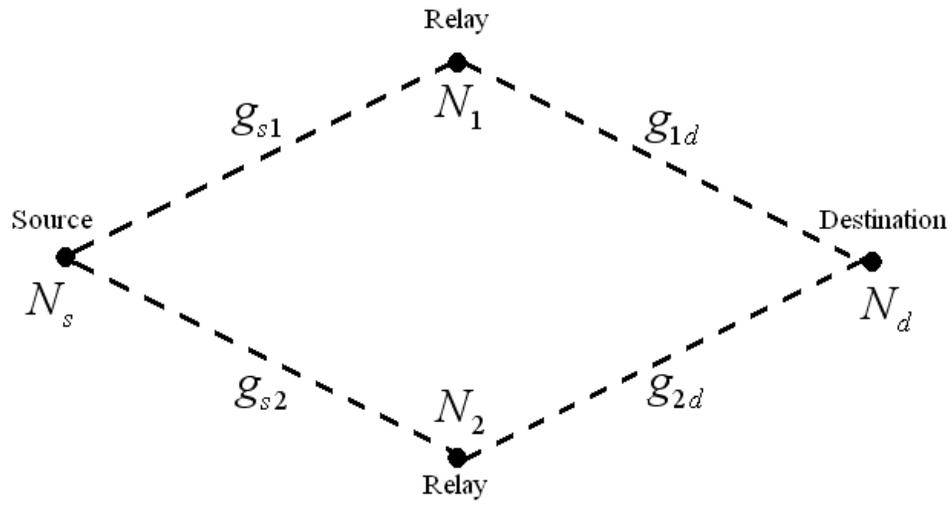}
\centering \caption{ Diamond relay network topology.} \label{fig_2}
\end{figure}

\begin{figure}[h]
\centering
\includegraphics[width = 5in]{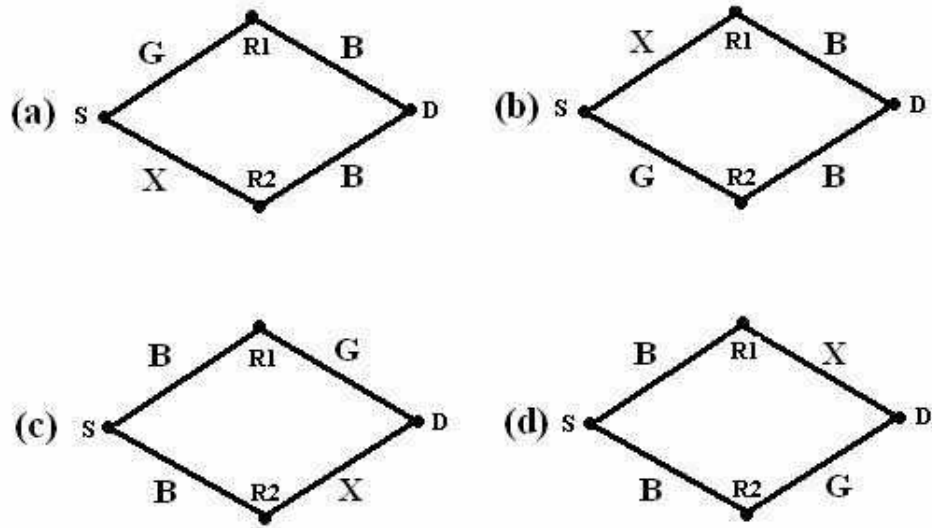}
\caption{ The case in which the new developed strategy with buffers
works.} \label{fig_3}
\end{figure}

\begin{figure}[h]
\centering
\includegraphics[width = 5in]{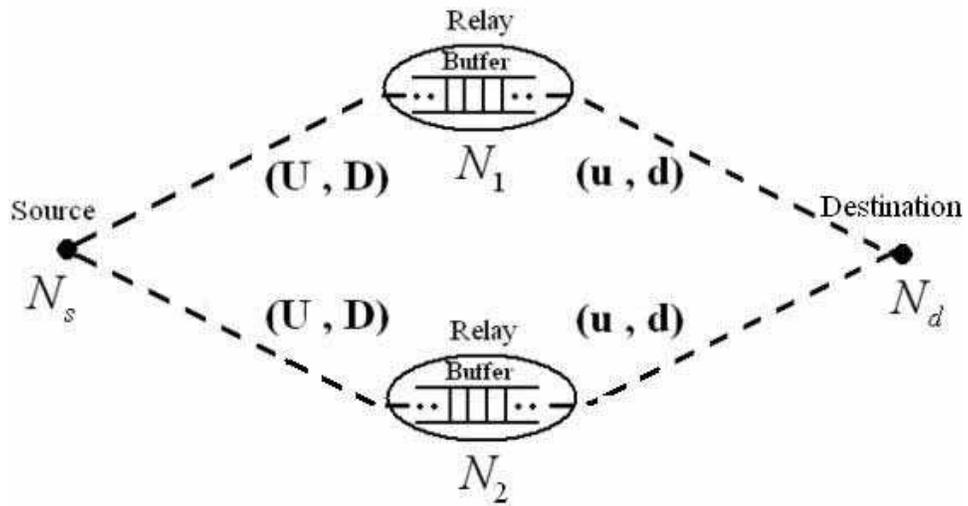}
\centering \caption{The model of $N_1$ and links marked with the
thresholds.} \label{fig_4}
\end{figure}

\begin{figure}[h]
\centering
\includegraphics[width = 5in]{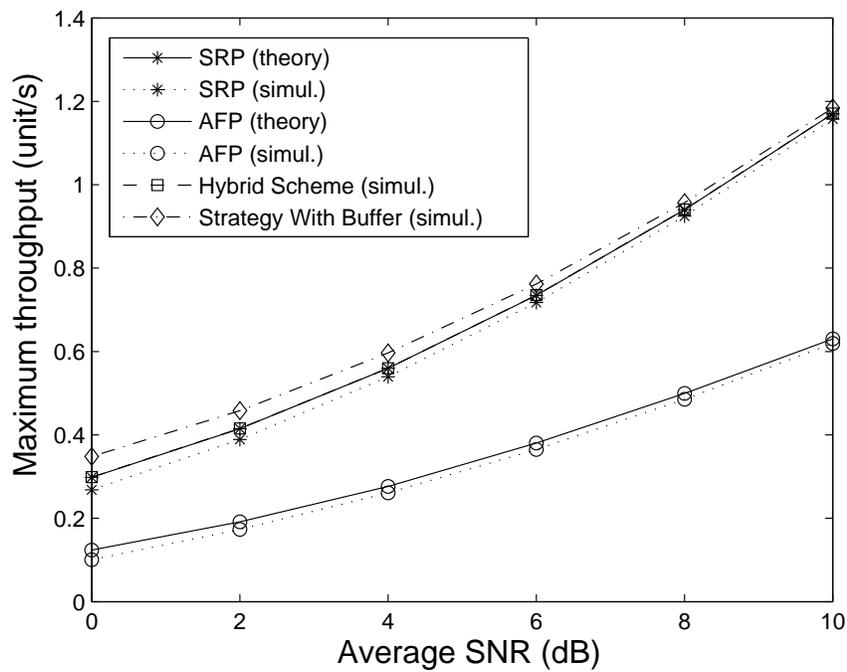}
\caption{The comparison of the throughput performance between
different schemes.} \label{fig_5}
\end{figure}

\begin{figure}[h]
\centering
\includegraphics[width = 5in]{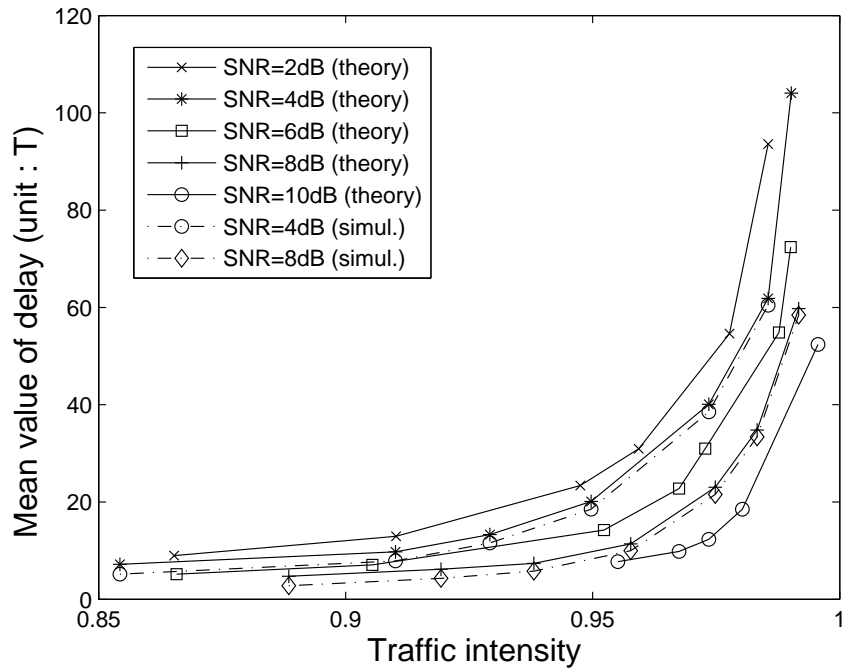}
\caption{The mean value of delay for different received SNR (T)
}\label{fig_6}
\end{figure}

\begin{figure}[h]
\centering
\includegraphics[width = 5in]{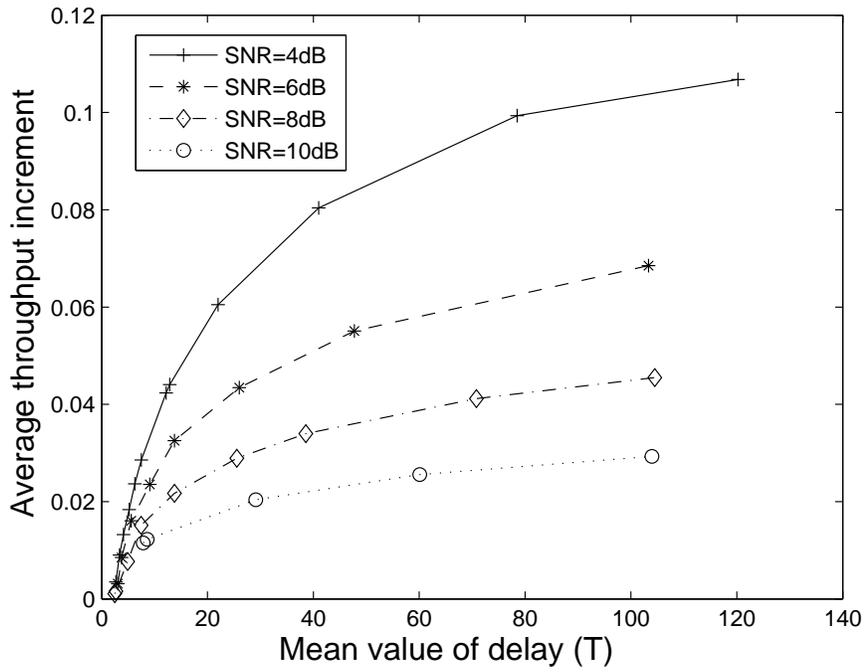}
\caption{ The tradeoff between the throughput improvement and the
average delay for different received SNR.} \label{fig_7}
\end{figure}

\begin{table}[h]
\tabcolsep 0pt \caption{maximum link-rates(unit/s) for different
$\overline{SNR}$ under 16-state levels} \vspace*{-33pt}
\begin{center}
\def\temptablewidth{0.5\textwidth}
{\rule{\temptablewidth}{1pt}}
\begin{tabular*}{\temptablewidth}{@{\extracolsep{\fill}}c|cccccccc}
$\overline{SNR}$(dB) &rank1 &rank2 &rank3 &rank4 &rank5 &rank6 &rank7 &rank8 \\
\hline
0&0.025 &0.07 &0.115 $\ast$&0.16 &0.205 &0.255 &0.305 &0.355\\
2&0.035 &0.105 &0.17 $\ast$&0.24 &0.32 &0.37 &0.435 &0.50\\
4&0.0505 &0.16 &0.255 $\ast$&0.35 &0.435 &0.52 &0.605 &0.685\\
6&0.085 &0.24 &0.375 $\ast$&0.495 &0.605 &0.71 &0.81 &0.91\\
8&0.13 &0.35 &0.525 $\ast$&0.68 &0.815 &0.935 &1.05 &1.16\\
10&0.20 &0.495 &0.715 $\ast$&0.90 &1.055 &1.19 &1.32 &1.435\\
\hline\hline
$\overline{SNR}$(dB) &rank9 &rank10 &rank11 &rank12 &rank13 &rank14 &rank15 &rank16 \\
\hline
0&0.405 &0.465 &0.525 &0.59 &0.67 &$\star$0.76 &0.88 &1.015\\
2&0.57 &0.64 &0.715 &0.795 &0.885 &$\star$0.99 &1.13 &1.28\\
4&0.77 &0.855 &0.94 &1.035 &1.135 &$\star$1.255 &1.40 &1.565\\
6&1.00 &1.10 &1.195 &1.30 &1.41 &$\star$1.535 &1.695 &1.865\\
8&1.265 &1.355 &1.475 &1.585 &1.705 &$\star$1.835 &2.005 &2.175\\
10&1.55 &1.66 &1.775 &1.89 &2.01 &$\star$2.15 &2.32 &2.495
       \end{tabular*}
       {\rule{\temptablewidth}{1pt}}
       \end{center}
       \end{table}

\begin{table}[h]
\tabcolsep 0pt \caption{notations defined for the analysis}
\vspace*{-3pt}
\begin{center}
\def\temptablewidth{0.5\textwidth}
{\rule{\temptablewidth}{1pt}}
\begin{tabular*}{\temptablewidth}{@{\extracolsep{\fill}}c|cc}
symbol & meanings &\\
\hline
$N$ & total number of the state levels &\\
$U$ & level number of $C_{upTH}$ for source-relay links &\\
$u$ & level number of $C_{upTH}$ for relay-destination links &\\
$D$ & level number of $C_{dwTH}$ for source-relay links &\\
$d$ & level number of $C_{dwTH}$ for relay-destination links &\\
$P_x$ & probability that at least one source-relay link is very good &\\
& while the two from relay-destination links are very bad &\\
$P_y$ & probability that at least one relay-destination link is very good &\\
& while the two from source-relay links are very bad &\\
$P$ & the sum of $P_x \,\,\, and \,\,\, P_y$ &\\
$a$ & arrival interval of the relay node\\
$b$ & service time of the relay node\\
$C_k$ & maximum rate of the link when its state is at level $k$ &\\
$t_U$ & set of arrival intervals when link state level is above or equal to U &\\
$t_D$ & set of arrival intervals when link state level is below or equal to D &\\
$t_u$ & set of service time when link state level is above or equal to u &\\
$t_d$ & set of service time when link state level is below or equal to d &\\
$E(a)$ & mean value of arrival interval &\\
$E(b)$ & mean value of service time &\\
$\sigma_a^2$ & variance of arrival interval &\\
$\sigma_b^2$ & variance of service time &\\
$\rho$ & traffic intensity of the buffer &\\
$\overline{W}$ & the mean value of delay &
       \end{tabular*}
       {\rule{\temptablewidth}{1pt}}
       \end{center}
       \end{table}

\begin{table}[h]
\tabcolsep 0pt \caption{delay, improvement of rate, $C_{upTH}$ and
$C_{dwTH}$ for $SNR=6dB$ (16 channel states)} \vspace*{-33pt}
\begin{center}
\def\temptablewidth{0.5\textwidth}
{\rule{\temptablewidth}{1pt}}
\begin{tabular*}{\temptablewidth}{@{\extracolsep{\fill}}c|ccccc}
& & SNR & $=$ & 6dB &\\
\hline
average delay &6.68 &15.08 &17.27 &29.38 &$\infty$ \\
rate-improvement &0.0199 &0.0269 &0.0311 &0.0465 &$-$\\
$D$ &3 &3 &3 &3 &2\\
$d$ &1 &1 &2 &2 &2\\
$U$ &16 &16 &16 &15 &15\\
$u$ &14 &15 &15 &15 &15
       \end{tabular*}
       {\rule{\temptablewidth}{1pt}}
       \end{center}
       \end{table}

\end{document}